# Complex Lattice Reduction Algorithm for Low-Complexity MIMO Detection

Ying Hung Gan, *Student Member, IEEE*, Cong Ling, *Member, IEEE*, and Wai Ho Mow, *Senior Member, IEEE*

*Abstract*— Recently, lattice-reduction-aided detectors have been proposed for multiple-input multiple-output (MIMO) systems to give performance with full diversity like maximum likelihood receiver, and yet with complexity similar to linear receivers. However, these lattice-reduction-aided detectors are based on the traditional LLL reduction algorithm that was originally introduced for reducing real lattice bases, in spite of the fact that the channel matrices are inherently complex-valued. In this paper, we introduce the complex LLL algorithm for direct application to reduce the basis of a complex lattice which is naturally defined by a complex-valued channel matrix. We prove that complex LLL reduction-aided detection can also achieve full diversity. Our analysis reveals that the new complex LLL algorithm can achieve a reduction in complexity of nearly 50% over the traditional LLL algorithm, and this is confirmed by simulation. It is noteworthy that the complex LLL algorithm aforementioned has nearly the same bit-error-rate performance as the traditional LLL algorithm.

## I. Introduction

**B**Y exploiting the linearity of a communication channel and the lattice structure of the modulation, many detection problems can be interpreted as the problem of finding the closest lattice point. This lattice viewpoint of detection problems [1]–[3] forms the foundation of many low-complexity high-performance lattice-based detectors, such as Pohst's sphere decoder and various approximate lattice decoders (see [4] and the references therein). However, since the traditional lattice formulation is only directly applicable to a real-valued channel matrix, most conventional lattice-based detectors were derived based on the real-valued channel matrix of the complex-valued channel matrix. This approach doubles the channel matrix dimension and may lead to an unnecessarily complicated detector. This insight suggests the possibility of deriving even simpler detectors for complex-valued channel matrices by introducing the *complex lattice* formulation.

Recently, by exploiting the lattice structure of wireless multiple-antennas systems, *lattice reduction* is employed to improve the performance of MIMO detection [4]–[8] and precoding [9]. The most commonly used and practical lattice reduction algorithm is the LLL reduction algorithm [10]. Its generalization, not just to the complex number field, but to the Euclidean ring in general, was introduced in [11]. In this paper, we investigate the implementation complexity and performance of of the complex LLL algorithm for the application of MIMO detection. From analytic and simulation results, it is shown that the average overall complexity of our accelerated complex LLL (CLLL) reduction algorithm is nearly half of that of the real LLL (RLLL) reduction algorithm. Like the RLLL algorithm, linear detectors employing the CLLL algorithm can achieve full diversity in lattice-reduction-aided decoding. Moreover, the bit-error-rate performance of MIMO detection schemes using CLLL-reduced basis is practically the same as that using RLLL-reduced basis. Thus, we achieve a reduction as large as 50% in the complexity of the reduction algorithm without sacrificing any performance.

LLL reduction is often treated as part of preprocessing and hence its complexity is shared by symbols within the coherence time. However, in the situation where the channel matrix changes relatively rapidly, i.e. fast fading channel, the complexity of this preprocessing part becomes crucial. Our reduction algorithm makes it feasible to use reduction-aided detection schemes, which have much better performance than traditional schemes, even in fast fading channel.

The rest of the paper is organized as follows. In Section II, the system model and its lattice viewpoint, along with the notations used throughout this paper, are given. The CLLL reduction algorithm and its complexity analysis are then described in Section III. Section IV proves the achievability of full diversity. In Section V, we present our simulation results. Finally, the paper is concluded in Section VI.

## II. Preliminaries

### A. System model

Consider an $n \times m$ MIMO system consists of $n$ transmitters and $m$ receivers. The relationship between the transmitted column vector $\mathbf{x}$ and the received vector $\mathbf{y}$ is determined by

$$\mathbf{y} = \mathbf{H}\mathbf{x} + \mathbf{w}, \qquad (1)$$

where $\mathbf{H} = [\mathbf{h}_1 \cdots \mathbf{h}_n]$ representing a flat-fading channel gain matrix, is a $m \times n$ complex matrix, all of its elements are independent complex Gaussian random variables $CN(0,1)$, and $\mathbf{w}$ is the additive noise vector, all of its elements are independent complex random variable $CN(0, 2\sigma^2)$. We assume a full-rank MIMO system, i.e. $\mathbf{H}$ consists of $n \ (\leq m)$ linearly independent vectors. The transmitted vector $\mathbf{x}$ is drawn from a finite set $\mathcal{C}$, representing the complex constellation being used.

Submitted to *IEEE Trans. Wireless Commun.* in March 2006. This work was presented in part at the 2005 Global Telecommunications Conference, United States, November 2005.

This work was supported by the Hong Kong Research Grants Council.

Y. H. Gan (yh.gan@ieee.org) and W. H. Mow (w.mow@ieee.org) are with the Department of Electrical and Electronic Engineering, Hong Kong University of Science and Technology, Clearwater Bay, Hong Kong.

C. Ling (cling@ieee.org) is with the Department of Electronic Engineering, King's College London, Strand, London WC2R 2LS, United Kingdom.

### B. The complex lattice viewpoint

A *complex lattice* $\Lambda$ is the set of points $\{\sum_{i=1}^{n} c_i \mathbf{h}_i : c_i \in \mathcal{G}, \mathbf{h}_i \in \mathbb{C}^m\}$, where $\mathcal{G}$ is the set of complex integers $\mathcal{G} = \mathbb{Z} + i\mathbb{Z}$, $i = \sqrt{-1}$. In the matrix form, $\Lambda = \{\mathbf{Hc} : \mathbf{c} \in \mathcal{G}^n\}$, where $\mathbf{H} = [\mathbf{h}_1 \cdots \mathbf{h}_n]$ represents a *basis* of the lattice $\Lambda^\dagger$. The lattice $\Lambda$ can have infinitely many different bases other than $\mathbf{H}$. In general, any matrix $\mathbf{H}'$ such that $\mathbf{H}' = \mathbf{HU}$, where $\mathbf{U}$ is an *unimodular* matrix (i.e. $|\det \mathbf{H}| = 1$ and all elements of $\mathbf{H}$ are complex integers), is also a basis of $\Lambda$.

A *lattice reduction algorithm* is an algorithm that, given $\mathbf{H}$, finds another basis $\mathbf{H}'$ which enjoys several "good" properties. There are many definitions of lattice reduction, such as Minkowski reduction [12] and Korkine-Zolotareff (KZ) reduction [13]. (See, e.g.[14]–[17] for a modern description of them.) Among them, the most practically used is the *LLL reduction* algorithm (named after its inventors Lenstra, Lenstra and Lovász [10]), whose running time is polynomial in the dimension of the lattice.

The LLL reduction can be employed to improve the performance of MIMO detection schemes, assuming that the channel state information (i.e. the matrix $\mathbf{H}$) is perfectly known at the receiver. Since the vector $\mathbf{Hx}$ can be viewed as a lattice point[2], MIMO detection can be formulated as finding a reasonably close lattice point to $\mathbf{Hx}$ given $\mathbf{y}$. By considering the reduced basis $\mathbf{H}'$ instead of $\mathbf{H}$, it can be shown that the performance of traditional detection schemes like *zero-forcing* (ZF) and *successive interference cancellation* (SIC) can be greatly improved [4], [5]. Consider the MIMO system

$$\mathbf{y} = \mathbf{Hx} + \mathbf{w} = \mathbf{H'U}^{-1}\mathbf{x} + \mathbf{w} = \mathbf{H'x'} + \mathbf{w}. \qquad (2)$$

For successive interference cancellation, we first apply QR decomposition to the reduced basis

$$\mathbf{H}' = \mathbf{QR}, \qquad (3)$$

such that $\mathbf{Q}$ is unitary and $\mathbf{R}$ is upper triangular. Then multiply the Hermitian of $\mathbf{Q}$ to $\mathbf{y}$: $\mathbf{Q}^H\mathbf{y}$ and employ successive decision and cancellation [18] to obtain a hard-decision vector $\hat{\mathbf{x}}'$. Apply the unimodular transform to $\hat{\mathbf{x}}'$ to obtain $\tilde{\mathbf{x}} = \mathbf{U}\hat{\mathbf{x}}'$ and finally, hard-limit $\tilde{\mathbf{x}}$ component-wise to a valid symbol vector $\hat{\mathbf{x}}$. Figure 1 shows the block diagram of such a system. The performance is much better than performing successive interference cancellation detection on the original basis $\mathbf{H}$. In fact, it has been proved that lattice-aided detection schemes using real lattice achieve full diversity [19], [20] . For more details on this lattice viewpoint, refer to [4], [5], [14].

### C. Notation

The following notation is used throughout this paper. Let $\bar{z}$ represent the complex conjugate of $z$. The conjugate transpose (Hermitian) of a matrix $\mathbf{H}$ is denoted by $\mathbf{H}^H$. Denote by $\mathbf{H}^\dagger$ the Moore-Penrose pseudo-inverse of $\mathbf{H}$. The inner product of two vectors $\mathbf{h}_1$ and $\mathbf{h}_2$ is defined as $\langle \mathbf{h}_1, \mathbf{h}_2 \rangle = \mathbf{h}_2^H \mathbf{h}_1$. The set of orthogonal vectors generated by the *Gram-Schmidt Orthogonalization (GSO)* procedure are represented as $\{\mathbf{h}_1^*, \ldots, \mathbf{h}_n^*\}$ which span the same space as $\{\mathbf{h}_1, \ldots, \mathbf{h}_n\}$, and further let $\mu_{ij} = \frac{\langle \mathbf{h}_i, \mathbf{h}_j^* \rangle}{\|\mathbf{h}_j^*\|^2}$. The squared norm of $\mathbf{h}_i^*$ is denoted as $\mathcal{H}_i = \|\mathbf{h}_i^*\|^2$. $\Re(z)$ and $\Im(z)$ denote the respective real and complex parts of a complex number $z$. Besides, $\lfloor z \rceil$ rounds $z$ to the nearest integer, and if $z$ is a complex number, it is done on both the real part and the complex part, i.e. $\lfloor z \rceil = \lfloor \Re(z) \rceil + i\lfloor \Im(z) \rceil$. For real number $r$, $|r|$ denotes its absolute value. For complex number $z$, $|z|$ denotes its modulus: $|z| = \sqrt{z\bar{z}}$.

### III. Complex LLL Reduction Algorithm

In this section we describe a CLLL reduction algorithm. Traditionally, LLL reduction is performed on the real-valued equivalent matrix of the complex matrix $\mathbf{H}$ [1, pp.81] (c.f. [8], [9]):

$$\mathbf{H}_R = \begin{bmatrix} \Re(\mathbf{H}) & -\Im(\mathbf{H}) \\ \Im(\mathbf{H}) & \Re(\mathbf{H}) \end{bmatrix}. \qquad (4)$$

Since the reduced basis matrix does not generally have the symmetric structure as in (4), the detection part also has to be done in the real number field (c.f. [8]). This conversion, which causes a doubling in dimension, generally requires extra computations in the detection algorithm (during the processing phase) as well as the lattice reduction algorithm (during the preprocessing phase).

Our algorithm works directly on complex lattices. Although the cost for each complex arithmetic operation is higher than its real counterpart, the total number of operations required for CLLL is fewer, leading to a lower overall complexity of CLLL. Moreover, the quality of CLLL-reduced basis is the same as that of the real LLL (RLLL) reduced basis.

### A. Principle of LLL reduction

A basis $\mathbf{H}$ for a complex lattice is *CLLL-reduced* if both of the following conditions are satisfied:

$$|\Re(\mu_{ij})| \le 0.5 \quad \text{and} \quad |\Im(\mu_{ij})| \le 0.5 \qquad (5)$$

for $1 \le j < i \le n$, and

$$\mathcal{H}_k \ge (\delta - |\mu_{k,k-1}|^2)\mathcal{H}_{k-1} \qquad (6)$$

for $1 < k \le n$, where $\delta$ with $1/2 < \delta < 1$ is a factor selected to achieve a good quality-complexity tradeoff [10].

Note that the value $\delta = 1$ can be used as well, but polynomial convergence of the algorithm is not guaranteed. If $n = 2$ and $\delta = 1$, then the basis has the properties $|\Re(\mu_{2,1})| \le 0.5$, $|\Im(\mu_{2,1})| \le 0.5$, and $\mathcal{H}_1 \le \mathcal{H}_2$. This is exactly the complex Gaussian reduction presented by Yao and Wornell [5]. For real lattices, it is well known that 2D RLLL reduction with $\delta = 1$ is equivalent to Gaussian reduction. We can see the parallel between real and complex lattice reductions.

The CLLL reduction algorithm consists of 3 steps:

1. A modified GSO procedure as in [4] to compute $\mathcal{H}_i$;
2. *Size reduction* process that aims to make basis vectors shorter and closer to orthogonal by asserting (5) for all $j < i$;

---

[1] Since the lattice basis is always arisen from the channel matrix, without the ambituigity, we use the same symbol $\mathbf{H}$ to denote both.

[2] To apply the lattice viewpoint, the constellation $\mathcal{C}$ has to be of lattice type, such as BPSK, QPSK, QAM or PAM.

3) *Basis vectors swapping* step. Two consecutive basis vectors $\mathbf{h}_{k-1}$ and $\mathbf{h}_k$ will be swapped if (6) is violated. The idea is that, after swapping, size reduction can be repeated to make basis vectors shorter.

The two steps, size reduction and basis vectors swapping, iterate until (6) is satisfied by all pairs of $\mathbf{h}_{k-1}$ and $\mathbf{h}_k$. The resultant basis is thus CLLL-reduced.

Algorithm 1 gives the detailed description of the CLLL reduction algorithm. The algorithm also computes the unimodular matrix $\mathbf{U}$, which is required for our lattice-reduction-aided detection. This saves computational cost over explicitly calculating $\mathbf{U} = (\mathbf{H}^{\dagger})\mathbf{H}'$.

It is noteworthy that our algorithm has three distinctions from Napias [11]:

- The condition (5) of size reduction is stronger than $|\mu_{ij}|^2 \leq 0.5$ given in [11], resulting in fewer size reduction computations.
- The algorithm checks if $|\Re(\mu_{k,k-1})| > 0.5$ or $|\Im(\mu_{k,k-1})| > 0.5$ before doing size reduction, whereas [11] does not. The check will avoid unnecessary computations and will improve the efficiency accordingly.
- After swapping $\mathbf{h}_{k-1}$ and $\mathbf{h}_k$, $\mathcal{H}_{k-1}$, $\mathcal{H}_k$ and some of the $\mu_{ij}$'s needed to reflect the change. Instead of calling the GSO procedure again as in [11], the following updating formulas are executed to eliminate any unnecessary operations

$$\dot{\mathbf{h}}_{k-1} = \mathbf{h}_k, \tag{7}$$

$$\dot{\mathbf{h}}_k = \mathbf{h}_{k-1}, \tag{8}$$

$$\dot{\mathcal{H}}_{k-1} = \mathcal{H}_k + |\mu_{k,k-1}|^2 \mathcal{H}_{k-1}, \tag{9}$$

$$\dot{\mu}_{k,k-1} = \overline{\mu_{k,k-1}} \left( \frac{\mathcal{H}_{k-1}}{\dot{\mathcal{H}}_{k-1}} \right), \tag{10}$$

$$\dot{\mathcal{H}}_k = \left( \frac{\mathcal{H}_{k-1}}{\dot{\mathcal{H}}_{k-1}} \right) \mathcal{H}_k, \tag{11}$$

$$\dot{\mu}_{i,k-1} = \mu_{i,k-1}\dot{\mu}_{k,k-1} + \mu_{ik}\frac{\mathcal{H}_k}{\dot{\mathcal{H}}_{k-1}}, \quad k < i \leq n, \tag{12}$$

$$\dot{\mu}_{ik} = \mu_{i,k-1} - \mu_{ik}\mu_{k,k-1}, \quad k < i \leq n, \tag{13}$$

$$\dot{\mu}_{k-1,j} = \mu_{kj}, \quad 1 \leq j \leq k-2, \tag{14}$$

$$\dot{\mu}_{k,j} = \mu_{k-1,j}, \quad 1 \leq j \leq k-2, \tag{15}$$

where $\dot{\mathbf{h}}_i$, $\dot{\mathcal{H}}_i$ and $\dot{\mu}_{ij}$ denote the updated values of $\mathbf{h}_i$, $\mathcal{H}_i$ and $\mu_{ij}$ respectively.

The same algorithm can also be employed to reduce a real-valued lattice basis without any modification. Hence it can be viewed as a generalization of the traditional LLL algorithm.

### B. Complexity analysis

The complexity of the LLL reduction algorithm depends on the distribution of the random basis matrix $\mathbf{H}$. For simplicity, we assume that $n = m$, i.e. $\mathbf{H}$ is a square matrix.

The speed of convergence can be determined by examining the product [10] (c.f. [4])

$$D = \prod_{i=1}^{n} d_i, \quad d_i = \prod_{j=1}^{i} \mathcal{H}_j. \tag{16}$$

---

**Algorithm 1.** CLLL Reduction Algorithm

*Input:* Lattice basis $\mathbf{H} = [\mathbf{h}_1 \cdots \mathbf{h}_n]$, factor $\delta \in (\frac{1}{2}, 1)$.
*Output:* CLLL-reduced basis $\mathbf{H}'$, unimodular matrix $\mathbf{U} = [\mathbf{u}_1 \cdots \mathbf{u}_n]$.

1: **for** $j = 1$ to $n$ **do**
2:    $\mathcal{H}_j \leftarrow \langle \mathbf{h}_j, \mathbf{h}_j \rangle$
3: **end for**
4: **for** $j = 1$ to $n$ **do**
5:    **for** $i = j + 1$ to $n$ **do**
6:       $\mu_{ij} \leftarrow \frac{1}{\mathcal{H}_j} \left( \langle \mathbf{h}_i, \mathbf{h}_j \rangle - \sum_{k=1}^{j-1} \overline{\mu_{jk}} \mu_{ik} \mathcal{H}_k \right)$
7:       $\mathcal{H}_i \leftarrow \mathcal{H}_i - |\mu_{ij}|^2 \mathcal{H}_j$
8:    **end for**
9: **end for**
10: $\mathbf{U} \leftarrow \mathbf{I}_n$
11: $k \leftarrow 2$
12: **while** $k \leq n$ **do**
13:    **if** $|\Re(\mu_{k,k-1})| > \frac{1}{2}$ or $|\Im(\mu_{k,k-1})| > \frac{1}{2}$ **then**
14:       $\{\mathbf{H}, \mu\} \leftarrow$ SIZE_REDUCE$(\mathbf{H}, \mu, k, k-1)$
15:    **end if**
16:    **if** condition (6) is violated on $k$ and $k-1$ **then**
17:       swap and update using formulas (7)-(15)
18:       swap $\mathbf{u}_k$ and $\mathbf{u}_{k-1}$
19:       $k \leftarrow \max(2, k-1)$
20:    **else**
21:       **for** $j = k-2$ to $1$ step $-1$ **do**
22:          **if** $|\Re(\mu_{kj})| > \frac{1}{2}$ or $|\Im(\mu_{kj})| > \frac{1}{2}$ **then**
23:             $\{\mathbf{H}, \mu\} \leftarrow$ SIZE_REDUCE$(\mathbf{H}, \mu, k, j)$
24:          **end if**
25:       **end for**
26:       $k \leftarrow k + 1$
27:    **end if**
28: **end while**
29: return $\mathbf{H}$ as $\mathbf{H}'$, and $\mathbf{U}$.

---

$D$ only changes when two basis vectors are swapped, i.e. $\mathcal{H}_k < (\delta - |\mu_{k,k-1}|^2)\mathcal{H}_{k-1}$. After swapping, $d_i$ contracts by a factor of $\delta$, while other $d_i$'s are unchanged. As a result, $D$ contracts by a factor of at least $\delta$. Now it is clear that the parameter $\delta$ dictates the speed of convergence. Hence, it is expected that CLLL and RLLL algorithms with the same value of $\delta$ have a similar speed of convergence.

Moreover, the norm $B_c$ of the longest column vector of $\mathbf{H}$ is equal to the norm $B_r$ of the longest column vector of $\mathbf{H}_R$, because

$$B_r = [\Re(\mathbf{h})]^2 + [\pm\Im(\mathbf{h})]^2 = B_c \tag{17}$$

which is clear from the form of $\mathbf{H}_R$ in (4). Define $B \triangleq B_r = B_c$, it can be shown that the number of basis swapping performed for any $\mathbf{H}$ is O$(n^2 \log B)$ [10] and the same applies to the case in which (6) is satisfied. Since the whole algorithm starts at $k = 2$, and must terminate when $k = n$, the part between lines 21-25 must be executed for O$(n^2 \log B) + n = $ O$(n^2 \log B)$ times. Therefore it can be easily seen that the complexity of LLL is actually dominated by this part, with overall complexity O$(n^4 \log B)$.

To preliminarily estimate how much complexity can be

Algorithm 2.   Subroutine SIZE_REDUCE

*Input:* $\mathbf{H} = [\mathbf{h}_1 \cdots \mathbf{h}_n]$, $\mu$, indices $k, j$, $1 \leq j < k \leq n$.
*Output:* $\mathbf{H}, \mu$.

1: $c \leftarrow \lfloor \mu_{kj} \rceil$
2: $\mathbf{h}_k \leftarrow \mathbf{h}_k - c\mathbf{h}_j$
3: $\mathbf{u}_k \leftarrow \mathbf{u}_k - c\mathbf{u}_j$
4: **for** $l = 1$ to $j$ **do**
5: $\quad \mu_{k,l} \leftarrow \mu_{k,l} - c\mu_{j,l}$
6: **end for**

saved by applying CLLL, compared with RLLL, we consider the *CLLL-to-RLLL Complexity Ratio* (CRCR)

$$K \frac{n^4 \log B}{(2n)^4 \log B}, \tag{18}$$

where $K$ is an architecture-dependent factor meaning, on average, how many real arithmetic operations have to be executed per each complex operation. For example, if a complex addition requires 2 real arithmetic operations and a complex multiplication requires 6, then $K = (6 + 2)/2 = 4$ since the number of additions and multiplications are the same.

However, there is one more factor affecting the complexity that we need to consider as well. Line 22 introduces a conditional test such that the execution of the subroutine SIZE_REDUCE at line 23 may be skipped sometimes. Denote $P_c$ ($P_r$) as the probability that this test is passed in CLLL (RLLL):

$$P_r(2n) = P\{|\mu_{k,k-1}(2n)| > 0.5\}, \quad \mu_{k,k-1}(2n) \text{ real} \tag{19}$$

and

$$P_c(n) = P\{|\Re[\mu_{k,k-1}(n)]| > 0.5 \text{ or } |\Im[\mu_{k,k-1}(n)]| > 0.5\}, \tag{20}$$

where, for clarity, the dependence on the dimension is shown explicitly. Then the CRCR (18) should be rewritten as:

$$K \frac{n^2 \log B \cdot P_c(n) \cdot n^2}{(2n)^2 \log B \cdot P_r(2n) \cdot (2n)^2} = \frac{K}{16} \frac{P_c(n)}{P_r(2n)}. \tag{21}$$

By definition of $\mu_{k,k-1}$, the random variables $\Re[\mu_{k,k-1}(n)], \Im[\mu_{k,k-1}(n)]$, and real-valued $\mu_{k,k-1}(2n)$ should have similar statistics. Moreover, result of our empirical studies, as shown in Table I for $n \leq 22$, indicated that both $P_r(2n)$ and $P_c(n)$ are small, decreasing with $n$. It is reasonable to assume that the event $|\Re(\mu_{k,k-1})| > 0.5$ and $|\Im(\mu_{k,k-1})| > 0.5$ are statistically independent for circular symmetric complex Gaussian $\mathbf{H}$. Then

$$
\begin{aligned}
P_c(n) &\approx P\{|\Re[\mu_{k,k-1}(n)]| > 0.5\} \\
&\quad + P\{|\Im[\mu_{k,k-1}(n)]| > 0.5\} \\
&= 2P\{|\Re[\mu_{k,k-1}(n)]| > 0.5\} \\
&= 2P_r(2n).
\end{aligned} \tag{22}
$$

Substituting this into (21) and using the common value $K = 4$, we have

$$\text{CRCR} \approx 4/16 \times 2 = 1/2 \tag{23}$$

which means that CLLL algorithm will have half of the complexity of RLLL algorithm. Empirical results to be presented will confirm the above prediction of 50% complexity reduction.

Finally, it is important to note the implication of the complex lattice approach on the complexity of other components in lattice-reduction-aided detectors. If ZF or SIC is used in MIMO detection, the computation of the pseudo-inverse or the QR decomposition of the reduced channel matrix should be considered as part of the preprocessing as well.

Since the RLLL-reduced basis matrix does not necessarily have a symmetric structure as (4), the pseudo-inverse or QR decomposition must be computed in the real number field. This, too, induces an extra complexity cost, as both pseudo-inverse and QR decomposition require $\mathrm{O}(n^3)$ field operations [21]. By employing the CLLL reduction, hence avoiding the doubling of dimension, only $K/2^3 = 1/2$ for $K = 4$ of operations are needed.

The processing part, on the other hand, requires $\mathrm{O}(n^2)$ field operations. When $K = 4$, the complexity reduction obtained by avoiding dimension doubling is cancelled by the extra number of flops required for complex arithmetics. If more complicated schemes which require $\Omega(n^{2+\xi})$, $\xi > 0$ field operations (e.g. V-BLAST [22]) are used, the complexity may also be reduced in the processing part by performing computations in in the complex number field.

## IV. FULL DIVERSITY

Computer simulation always sees that lattice-reduction-aided detection achieves the full diversity of a MIMO fading channel (e.g. [4], [5], [8], [9]). For the case of real lattice and $\delta = 3/4$, the achievability has been proved in [19]. [20, Appendix] not only showed the full diversity, but also determined the gap between maximum likelihood detection (MLD) and LLL reduction-aided decoding. A systematic approach was developed in [23] to obtain better upper bounds on the gap. To this end, [23] introduced proximity factors to measure the performance gap to MLD. It was shown that there exist constant upper bounds on the proximity factors for RLLL reduction. In this Section, we shall extend the analysis to CLLL reduction.

Consider a fixed but arbitrary $n$-D complex lattice $\Lambda$. The decision regions of ZF and SIC have $2n$ faces. We only have to study $n$ distances due to symmetry. The $i$-th distance of ZF is $d_{i,\text{ZF}} = ||\mathbf{h}_i|| \sin \theta_i$, for $i = 1, \ldots, n$, where $\theta_i$ denotes the acute angle between $\mathbf{h}_i$ and the linear space spanned by the other $n - 1$ basis vectors $\mathbf{h}_1, \ldots, \mathbf{h}_{i-1}, \mathbf{h}_{i+1}, \ldots, \mathbf{h}_n$. For the SIC detector, the $i$-th distance is given by $||\mathbf{h}_i^*||$.

If the boundary effects are ignored, the minimum distance of ML detection is obviously $d_{\text{ML}} = \lambda(\Lambda)$, where $\lambda(\Lambda)$ denotes the length of the shortest vector in $\Lambda$. Following [23], we define the *proximity factors* for CLLL reduction as

$$\rho_{i,\text{ZF}}^C \triangleq \sup \frac{\lambda^2(\Lambda)}{||\mathbf{h}_i||^2 \sin^2 \theta_i}, \tag{24}$$

$$\rho_{i,\text{SIC}}^C \triangleq \sup \frac{\lambda^2(\Lambda)}{||\mathbf{h}_i^*||^2} = \sup \frac{\lambda^2(\Lambda)}{\mathcal{H}_i}, \tag{25}$$

where the supremum is taken over the CLLL-reduced bases $\mathbf{H}$ of all $n$-D complex lattices. Furthermore, we define $\rho_{\mathrm{ZF}}^C \triangleq \max_{1 \le i \le n} \rho_{i,\mathrm{ZF}}^C$ and $\rho_{\mathrm{SIC}}^C \triangleq \max_{1 \le i \le n} \rho_{i,\mathrm{SIC}}^C$, which quantify the worst-case loss in the minimum squared Euclidean distance.

Using a union-bound argument [23], the error probability of ZF detection can be bounded as

$$P_{e,\mathrm{ZF}}(\mathrm{SNR}) \le \sum_{i=1}^{n} P_{e,\mathrm{ML}}\left(\frac{\mathrm{SNR}}{\rho_{i,\mathrm{ZF}}^C}\right) \le n P_{e,\mathrm{ML}}\left(\frac{\mathrm{SNR}}{\rho_{\mathrm{ZF}}^C}\right) \quad (26)$$

for arbitrary SNR and $\Lambda$. A similar bound exists for SIC detection if boundary errors are ignored. The relation (26) remains valid after averaging over $\mathbf{H}$. Obviously the error rate curve of lattice-reduction-aided detection only show a shift in SNR, up to a multiplicative factor $n$. It is clear that it can achieve full diversity, since we know MLD achieve full diversity. The proximity factors measure the performance gap between MLD and lattice-reduction-aided detection at high SNR.

In the following, we derive upper bounds on the proximity factors for CLLL reduction. The bounds are not necessarily the tightest possible; the main purpose here is to show the existence of constant bounds so that full diversity can be proven.

### A. CLLL-SIC

By definition (6) we have

$$\mathcal{H}_i \ge (\delta - |\mu_{i,i-1}|^2)\mathcal{H}_{i-1} \ge (\delta - 1/2)\mathcal{H}_{i-1}, \quad (27)$$

as $|\mu_{i,i-1}|^2 \le 1/2$ for CLLL reduction. By induction we can see

$$\mathcal{H}_j \le \alpha^{i-j}\mathcal{H}_i, \quad 1 \le j < i \le n, \quad (28)$$

where $\alpha \triangleq 1/(\delta - 1/2) \ge 2$. Substituting this into

$$\|\mathbf{h}_i\|^2 = \mathcal{H}_i + \sum_{j=1}^{i-1} |\mu_{ij}|^2 \mathcal{H}_j \quad (29)$$

due to the Gram-Schmidt orthogonalization, we obtain

$$\|\mathbf{h}_i\|^2 \le \left(1 + \sum_{j=1}^{i-1} \alpha^{i-j}/2\right) \mathcal{H}_i \quad (30)$$

$$= \left(1 + \frac{1}{2}\frac{\alpha^i - \alpha}{\alpha - 1}\right)\mathcal{H}_i. \quad (31)$$

Replacing $i$ by $j$ and substituting (28), we have

$$\|\mathbf{h}_j\|^2 \le \left(1 + \frac{1}{2}\frac{\alpha^j - \alpha}{\alpha - 1}\right)\alpha^{i-j}\mathcal{H}_i \quad (32)$$

for $1 \le j \le i \le n$. Since $\lambda(\Lambda) \le \|\mathbf{h}_j\|$ and (32) holds for all $j \le i$, the loss in the $i$-th squared Euclidean distance is bounded by

$$\rho_{i,\mathrm{SIC}}^C \le \min_{1 \le j \le i}\left(1 + \frac{1}{2}\frac{\alpha^j - \alpha}{\alpha - 1}\right)\alpha^{i-j}. \quad (33)$$

To obtain an exponential upper bound, we show in the Appendix that

$$\|\mathbf{h}_j\|^2 \le \alpha^{i-1}\mathcal{H}_i, \quad 1 \le j \le i \le n \quad (34)$$

for $\alpha \ge 2$. Note that (33) is a better upper bound in general, though.

The inequality (34) implies the bound

$$\rho_{i,\mathrm{SIC}}^C \le \alpha^{i-1} \quad (35)$$

and

$$\rho_{\mathrm{SIC}}^C \le \alpha^{n-1}. \quad (36)$$

In particular, if $n = 2$ and $\delta = 1$ ($\alpha = 2$), then $\rho_{\mathrm{SIC}}^C \le 2$. This agrees with the maximum loss of 3 dB for the 2-D complex Gaussian reduction of Yao and Wornell [5].

Let us compare with the $2n$-D RLLL reduction [23]

$$\rho_{\mathrm{SIC}}^R \le \beta^{2n-1}, \quad \beta = (\delta - 1/4)^{-1}. \quad (37)$$

If $-1$ in the exponent of $\alpha$ and $\beta$ in (36) and (37) is ignored, the ratio of the two factors is

$$\frac{\rho_{\mathrm{SIC}}^R}{\rho_{\mathrm{SIC}}^C} \approx \frac{\beta^{2n}}{\alpha^n} = \frac{(\delta - 1/2)^n}{(\delta - 1/4)^{2n}}. \quad (38)$$

For the common choice $\delta = 3/4$, we have

$$\frac{\rho_{\mathrm{SIC}}^R}{\rho_{\mathrm{SIC}}^C} \approx \frac{(1/4)^n}{(1/2)^{2n}} = 1, \quad (39)$$

which means RLLL and CLLL reduction have equal proximity factors, regardless of the dimension $n$. For any other value of $\delta$, we have $\rho_{\mathrm{SIC}}^C \ge \rho_{\mathrm{SIC}}^R$, because

$$(\delta - 1/4)^2 \ge \delta - 1/2 \quad (40)$$

where the equality holds if and only if $\delta = 3/4$.

### B. CLLL-ZF

We need the following lemma to establish the upper bound for ZF.

*Lemma 1:* For CLLL-reduced basis,

$$\sin \theta_i \ge \left(\frac{2}{2 + \sqrt{2}}\right)^{n-i} (\sqrt{\alpha})^{-n+1}. \quad (41)$$

Lemma 1 extends Babai's lower bound on $\theta_i$ [24] to CLLL reduction. The proof basically follows Babai and is given in the Appendix. Lemma 1 along with the trivial inequality $\|\mathbf{h}_i\| \ge \lambda(\Lambda)$ leads to

$$\rho_{i,\mathrm{ZF}}^C \le \sup \frac{1}{\sin^2 \theta_i} \le \left(\frac{3 + 2\sqrt{2}}{2}\right)^{n-i} \alpha^{n-1}. \quad (42)$$

Since the upper bound is maximized when $i = 1$, we have

$$\rho_{\mathrm{ZF}}^C \le \left(\frac{3 + 2\sqrt{2}}{2}\alpha\right)^{n-1}. \quad (43)$$

Comparing (43) and (36), we see that the constant $(3 + 2\sqrt{2})/2$ represents the inferiority of ZF to SIC.

When $n = 2$ and $\delta = 1$, $\sin \theta_1 \ge 2/(2 + \sqrt{2})$ and $\sin \theta_2 \ge \sqrt{2}/2$. However, it is quite obvious that $\theta_1$ must be equal to $\theta_2$ here (this implies that the lower bound on $\theta_1$ is not tight.) Hence $\rho_{\mathrm{ZF}}^C \le 2$, which again agrees with the 3 dB loss of complex Gaussian reduction [5].

Let us compare with the $2n$-D RLLL reduction [23]

$$\rho_{\text{ZF}}^R \leq \left(\frac{9\beta}{4}\right)^{2n-1}. \tag{44}$$

Again, ignore $-1$ in the exponent of $\alpha$ and $\beta$ in (43) and (44), the ratio is

$$\begin{aligned}
\frac{\rho_{\text{ZF}}^R}{\rho_{\text{ZF}}^C} &\approx \left[\frac{(9/4)^2}{(3+2\sqrt{2})/2}\frac{(\delta-1/2)}{(\delta-1/4)^2}\right]^n \\
&= \left[1.74 \times \frac{(\delta-1/2)}{(\delta-1/4)^2}\right]^n. \tag{45}
\end{aligned}$$

The right-hand side of (45) is greater than one if $\delta > 0.58$, and is less than one otherwise.

### C. Properties of CLLL-reduced basis

Without proof we give other properties of CLLL-reduced basis which are similar to those of RLLL-reduced basis [10]:

$$d(\Lambda) \leq \prod_{i=1}^{n} ||\mathbf{h}_i|| \leq \alpha^{n(n-1)/4} d(\Lambda), \tag{46}$$

$$||\mathbf{h}_1|| \leq \alpha^{(n-1)/4} d(\Lambda)^{1/n}, \tag{47}$$

$$\alpha^{1-n}\lambda_i \leq ||\mathbf{h}_i||^2 \leq \alpha^{n-1}\lambda_i, \tag{48}$$

where $d(\Lambda) \triangleq \sqrt{\det(\mathbf{H}^H \mathbf{H})}$ for a basis $\mathbf{H}$ of lattice $\Lambda$, and $\lambda_i$ is the $i$-th successive minimum of $\Lambda$. These properties show in various senses that the vectors of a CLLL-reduced basis are not too long.

## V. SIMULATION RESULTS

In this section we compare the average complexity and the error-rate performance when the reduced bases were used in MIMO detection.

The average complexity was measured in terms of the average number of floating-point operations (flops) used. Simulations were performed in Matlab, in which the number of flops equals 2 for complex addition and 6 for multiplication. For real numbers, both addition and multiplication require 1 flop. Moreover, we assumed the costs of rounding and hard-limiting are negligible when compared to floating-point addition and multiplication. The LLL factor $\delta$ was set to 0.99 in all cases for the best performance.

We demonstrate the average complexity of LLL-reduction-aided successive interference cancellation (LLL-SIC) detection scheme. The whole detection process can be divided into two parts: preprocessing and processing. The preprocessing part includes these operations:

- Lattice reduction of channel matrix $\mathbf{H}$.
- QR decomposition of the reduced channel matrix $\mathbf{H}'$.

And the processing part includes:

- Matrix multiplication $\mathbf{Q}^H \mathbf{y}$.
- Successive nulling and cancellation.
- Unimodular transformation $\mathbf{U}\hat{\mathbf{x}}'$, where $\hat{\mathbf{x}}'$ is the vector obtained by successive nulling and cancellation; and hard-limiting of the resultant vector to a valid modulation symbol vector.

Note that QR decomposition can be replaced by GSO, which decompose $\mathbf{H}'$ into $\mathbf{G}$ consists of orthogonal column vectors and upper triangular matrix $\mathbf{M}^T$ of all $\mu_{ij}$. One might wonder if some speed could be gained by actually keeping the whole vectors $\mathbf{h}_i^*$ (instead of just their squared norms $\mathcal{H}_i$) in our LLL reduction algorithm such that $\mathbf{G}$ could also be obtained after the reduction. However, by doing so, many extra flops are required for updating after basis vectors swapping. Asymptotically, the complexity of the LLL algorithm would be $O(n^5 \log B)$ instead of $O(n^4 \log B)$. Our simulation result also shows that, even in low dimensional systems, this "integrated" approach costs more than computing a "standalone" QR decomposition on the reduced basis afterward. Therefore we abandon this integrated approach.

Table II shows the average complexity of the preprocessing and processing part of LLL-SIC. It can be seen that, by working on the complex lattice, the complexity of the entire preprocessing part was reduced by 45.1% for $n = 2$ (i.e. a 2-transmitter-2-receiver system), and reduced by somewhere between 42.4% to 49.2% for larger $n$.

In particular, the complexity reduction of the proposed CLLL reduction algorithm over the traditional RLLL is about 44.2% to 50.5% for our selected range of $n$.

About 40.4% to 47.1% of the computation was saved by computing QR decomposition in the complex number field. If we also assume the number of complex additions are roughly the same as the number of complex multiplications for this part, 4 flops are required for each complex number operation on average. Thus, we expect the complexity reduction of this part approaches $4/8 = 50\%$ for sufficiently large $n$.

The vector-error-rate (VER) and bit-error-rate (BER) performance when traditional RLLL-reduced and CLLL-reduced basis are used in MIMO detection are shown in Figures 2 and 3 respectively. The VER is the probability that at least one symbol in the transmitted vector are incorrectly detected. The MIMO system under consideration is a 6-transmitter-6-receiver uncoded system using 64-QAM. The system configuration is to maximize the multiplexing gain. Namely, each transmitter transmits its own symbol stream, independent other transmitters' data streams. The lattice-reduction-aided detection schemes being examined are SIC and ZF. For the sake of comparison, the performance of ZF, SIC, V-BLAST and MLD detectors are also shown.

Both Figures 2 and 3 show that the CLLL and RLLL reduction algorithms result in practically identical VER and BER performance in MIMO detection, although the bounding values of their respective proximity factors are unequal in general. This can be explained as follows.

- The proximity factors only quantify the worst-case losses, whereas the average error rate matters in fading channels. The worst case may occur with very low probability so that the average loss can be similar.
- The bounds on the proximity factors are likely to be not tight enough. It is well known that the LLL reduction performs better than the theoretic exponential bound in practice.

Nonetheless, it is clear that full diversity of the detectors is correctly predicted by the analysis in Section IV.

## VI. Concluding Remarks

In this paper, we extended the traditional LLL algorithm for reducing complex lattices. The resultant complex LLL algorithm was applied to complex-lattice-reduction-aided MIMO detection. We derived constant upper bounds on the proximity factors, thereby proving the achievability of full diversity. We showed that the complexity of the complex LLL algorithm is nearly half of that of the traditional algorithm. The complex LLL algorithm can achieve an average complexity saving of nearly 50% with negligible performance loss. Apart from MIMO detection, CLLL reduction can also find applications in the design of complex lattice codes [25], [26].

## Appendix

### A. Proof of (34)

We need to prove

$$\left(1 + \frac{1}{2}\frac{\alpha^j - \alpha}{\alpha - 1}\right) \leq \alpha^{j-1}, \tag{49}$$

or equivalently

$$\begin{aligned}
\alpha^j - \alpha^{j-1} &\geq \alpha - 1 + \frac{1}{2}(\alpha^j - \alpha) \\
&= \frac{1}{2}\alpha^j + \frac{1}{2}\alpha - 1
\end{aligned} \tag{50}$$

for $\alpha \geq 2$. The inequality holds obviously when $j = 1$. We prove the rest by induction. Suppose that it holds when $j = k$; when $j = k + 1$

$$\begin{aligned}
\alpha^{k+1} - \alpha^k &= \alpha(\alpha^k - \alpha^{k-1}) \\
&\geq \alpha\left(\frac{1}{2}\alpha^k + \frac{1}{2}\alpha - 1\right) \\
&= \frac{1}{2}\alpha^{k+1} + \alpha\left(\frac{1}{2}\alpha - 1\right) \\
&\geq \frac{1}{2}\alpha^{k+1} + \frac{1}{2}\alpha - 1
\end{aligned}$$

where the second inequality follows from $\alpha \geq 2$ and $\alpha/2 - 1 \geq 0$. $\qquad\square$

### B. Proof of (41)

Following Babai's method [24], let $a_i = -1$ and $a_t$ for $t \neq i, 1 \leq t \leq n$ be arbitrary complex numbers, and define

$$\gamma_j = \sum_{t=j}^n a_t \mu_{tj} = a_j + \sum_{t=j+1}^n a_t \mu_{tj}. \tag{51}$$

We shall prove that

$$\sum_{j=i}^n |\gamma_j|^2 \geq \left(\frac{2}{2+\sqrt{2}}\right)^{2(n-i)} \tag{52}$$

if $|\mu_{lj}| \leq \sqrt{2}/2$ for $1 \leq j < l \leq n$. This can be proved by assuming the contrary, which leads to

$$|\gamma_j| < \varepsilon \triangleq \left(\frac{2}{2+\sqrt{2}}\right)^{n-i}, \quad i \leq j \leq n. \tag{53}$$

This in turn requires

$$|a_j| < \varepsilon \left(\frac{2+\sqrt{2}}{2}\right)^{n-j}, \quad i \leq j \leq n. \tag{54}$$

This is valid for $j = n$, since $|a_n| = |\gamma_n| < \varepsilon$ by (51). For $j < n$ we have

$$|a_j| = \left|\gamma_j - \sum_{t=j+1}^n a_t \mu_{tj}\right| \leq |\gamma_j| + \sum_{t=j+1}^n |a_t|\sqrt{2}/2. \tag{55}$$

Using reverse induction,

$$\begin{aligned}
|a_j| &< \varepsilon + \sum_{t=j+1}^n \varepsilon\left(\frac{2+\sqrt{2}}{2}\right)^{n-t}\frac{\sqrt{2}}{2} \\
&= \varepsilon\left(\frac{2+\sqrt{2}}{2}\right)^{n-j}.
\end{aligned} \tag{56}$$

Hence (54) is proven. Letting $j = i$, we have $|a_i| < 1$, but this contradicts $a_i = -1$. Thus (52) must be valid.

Once again, following Babai [24] we have

$$\sin^2\theta_i \geq \sum_{j=1}^n |\gamma_j|^2 \frac{\mathcal{H}_j}{||\mathbf{h}_i||^2}. \tag{57}$$

Then, excluding $j < i$ terms from (57) yields a lower bound

$$\sin^2\theta_i \geq \sum_{j=i}^n |\gamma_j|^2 \frac{\mathcal{H}_j}{||\mathbf{h}_i||^2}. \tag{58}$$

Substituting (28) and (34) yields,

$$\begin{aligned}
\sin^2\theta_i &\geq \frac{\sum_{j=i}^n |\gamma_j|^2 \alpha^{i-j} \cdot \mathcal{H}_i}{\alpha^{i-1} \cdot \mathcal{H}_i} \\
&\geq \frac{\left(\frac{2}{2+\sqrt{2}}\right)^{2(n-i)} \alpha^{i-n}}{\alpha^{i-1}} \\
&= \alpha^{1-n}\left(\frac{2}{2+\sqrt{2}}\right)^{2(n-i)},
\end{aligned} \tag{59}$$

and this completes the proof of (41). $\qquad\blacksquare$

TABLE I

PROBABILITY THAT THE CONDITIONAL TEST IN LINE 22 IS PASSED IN

CLLL ($P_c$) AND RLLL ($P_r$).

| $n$ | $P_c(n)$ | $P_r(2n)$ | $P_c/P_r$ |
|-----|----------|-----------|-----------|
| 4 | 0.5232 | 0.2214 | 2.3635 |
| 6 | 0.4401 | 0.1890 | 2.3288 |
| 8 | 0.3953 | 0.1755 | 2.2524 |
| 10 | 0.2898 | 0.1228 | 2.3604 |
| 12 | 0.2686 | 0.1185 | 2.2661 |
| 14 | 0.2463 | 0.1128 | 2.1839 |
| 16 | 0.2284 | 0.1076 | 2.1225 |
| 18 | 0.2103 | 0.1020 | 2.0606 |
| 20 | 0.1944 | 0.0962 | 2.0220 |
| 22 | 0.1782 | 0.0898 | 1.9839 |

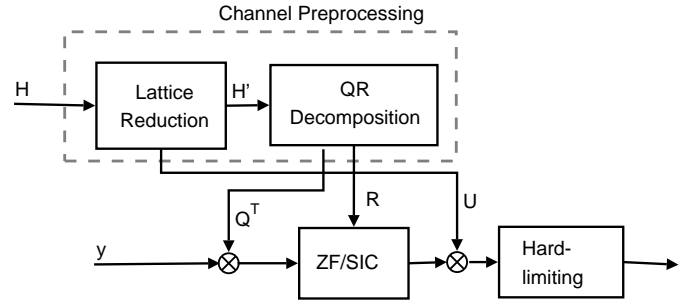

Fig. 1. Block diagram of the lattice-reduction-aided detectors.

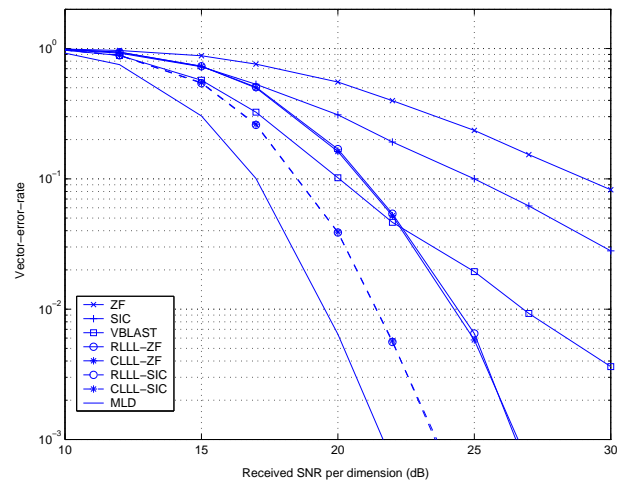

Fig. 2. The VER performance of various LLL reduction-aided MIMO detectors in a $6 \times 6$ uncoded MIMO system using 64-QAM.

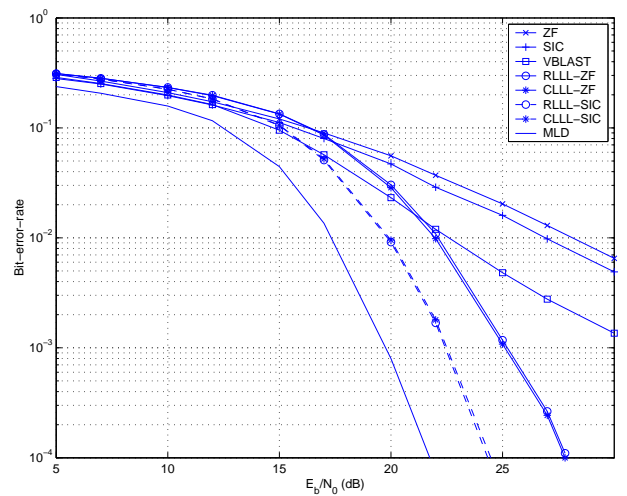

Fig. 3. The BER performance of various LLL reduction-aided MIMO detectors in a $6 \times 6$ uncoded MIMO system using 64-QAM.

TABLE II

SMALL CAPS: AVERAGE COMPLEXITY (IN FLOPS), ASSUMING 2 FLOPS PER COMPLEX ADDITION AND 6 FLOPS PER COMPLEX MULTIPLICATION.

| $n$ | Lattice reduction ($\delta = 0.99$) | | | QR decomposition: $\mathbf{QR} \leftarrow \mathbf{H}'$ | | | Overall % |
| | RLLL | CLLL | % saved | Real | Complex | % saved | saved |
|---|---|---|---|---|---|---|---|
| 2 | 275.29 | 145.05 | 47.3% | 273 | 156 | 42.9% | 45.1% |
| 3 | 979.14 | 546.00 | 44.2% | 845 | 504 | 40.4% | 42.4% |
| 4 | 2370.75 | 1351.56 | 43.0% | 1897 | 1116 | 41.2% | 42.2% |
| 6 | 8484.71 | 4787.58 | 43.6% | 6017 | 3420 | 43.2% | 43.4% |
| 8 | 21038.71 | 11557.68 | 45.1% | 13785 | 7644 | 44.6% | 44.9% |
| 10 | 42415.11 | 22524.70 | 46.9% | 26353 | 14364 | 45.5% | 46.4% |
| 12 | 73976.54 | 38387.97 | 48.1% | 44873 | 24156 | 46.2% | 47.4% |
| 14 | 118243.35 | 59788.66 | 49.4% | 70497 | 37596 | 46.8% | 48.4% |
| 16 | 176326.09 | 87264.24 | 50.5% | 104377 | 55260 | 47.1% | 49.2% |